\documentclass[
 aps,prb,
 long, numerical bibliography, (default)
numerical bibliography,
jmp,%
amsmath,amssymb,
reprint,superscriptaddress,
author-year,
author-numerical,
]{revtex4-1}

\usepackage{graphicx}
\usepackage{dcolumn}
\usepackage{bm}
\usepackage{xr}
\usepackage{epstopdf}
\begin{document}

\title{Temperature dependent Raman study of phonons of different symmetries in single crystal Bi$_2$Se$_3$}
\author{Bushra Irfan}
\affiliation{Department of Physics, Indian Institute of Technology (IIT), New Delhi 110016, India.}
\author{Satyaprakash Sahoo}
\affiliation{Department of Physics and Institute for Functional Nanomaterials, University of Puerto Rico, San Juan, PR00931USA.}
\author{Anand P. S. Gaur}
\affiliation{Department of Physics and Institute for Functional Nanomaterials, University of Puerto Rico, San Juan, PR00931USA.}
\author{Majid Ahmadi}
\affiliation{Department of Physics and Institute for Functional Nanomaterials, University of Puerto Rico, San Juan, PR00931USA.}
\author{Maxime J-F Guinel}
\affiliation{Department of Physics and Institute for Functional Nanomaterials, University of Puerto Rico, San Juan, PR00931USA.}
\affiliation{Department of Chemistry, College of Natural Sciences, University of Puerto Rico, PO Box 70377, San Juan, Puerto Rico 00936-8377, USA.}
\author{Ram S. Katiyar}
\affiliation{Department of Physics and Institute for Functional Nanomaterials, University of Puerto Rico, San Juan, PR00931USA.}
\author{Ratnamala Chatterjee}
\email{rmala@physics.iitd.ac.in}
\affiliation{Department of Physics, Indian Institute of Technology (IIT), New Delhi 110016, India.}

\date{\today}

\begin{abstract}
\textbf{High quality single crystals of Bi$_2$Se$_3$ were grown using a modified Bridgman technique, the detailed study were carried out using Raman spectroscopy and characterized by Laue diffraction and high resolution transmission electron microscopy. Polarized Raman scattering measurements were also carried out, and both the $A^1_g$ and $A^2_g$ phonon modes show strong polarization effect, which is consistent with the theoretical prediction. The temperature dependent study (in the temperature range  $83$ $K \le T \le 523$ $K$ ) of Raman active modes were reported and observed to follow a systematic red shift. The frequency of these phonon modes are found to vary linearly with temperature and can be explained by first order temperature co-efficient. The temperature co-efficient for $A^1_{1g}$, $E^2_g$ and $A^2_{1g}$ modes were estimated to be $-1.44\times10^{-2}$, $-1.94\times10^{-2}$ and $-1.95\times10^{-2}cm^{-1}/K$ respectively.}
\end{abstract}

\maketitle
\section{Introduction}
Bismuth selenide (Bi$_2$Se$_3$) is known to be a good thermoelectric material, however recently it has attracted much more attention and emerged as a new class of material known as topological insulators \cite{moore_birth_2010, fu_topological_2007}. Topological Insulators (TI) is a novel electronic state of matter with bulk insulating and topologically protected surface states in three dimension (3D) \cite{zhang_topological_2009}; and correspondingly edge state in two dimension \cite{qi_quantum_2010}. The interest in topological insulators is growing rapidly not only due to their exciting fundamental properties \cite{fu_superconducting_2008, chen_experimental_2009}, but also for their future applications in spintronics \cite{pesin_spintronics_2012} and quantum computing \cite{hasan_colloquium:_2010}. Understanding the behavior of quantum states in $3$D topological insulators is significant and one of the major challenges for carrier conduction in bulk. As phonon interacts strongly with charge carriers at room temperature a good knowledge of the vibrational properties of this material is essential to understand the carrier transport properties \cite{richter_raman_1977, cusco_temperature_2007}. Raman scattering is the fast and non-destructive technique that has been widely used to study electron-phonon interaction, phonon dynamics both in bulk and nano materials \cite{gouadec_raman_2007}. The band structure of Bi$_2$Se$_3$ surface states is very much similar to that of graphene \cite{castro_neto_electronic_2009, Hsieh_Observation_2009} and bismuth selenide also has a layered structure (as that of graphene) with quintuple layer of atoms stacked along c-axis. The five individual layer forming the quintuple layer occur as Se(1)-Bi-Se(2)-Bi-Se(1) and these quintuple layer coupled to another layer with weak Van der Waals forces \cite{mishra_electronic_1997}. Since the Raman peak position is dependent on the number of layers, it can be used to determine the number of layers in TIs \cite{calizo_temperature_2007}. Micro-Raman spectroscopy emerged as a valuable tool to study the number of atomic planes in two dimensional layer materials such as graphene \cite{ferrari_raman_2013, sahoo_polarized_2011}.

Also, a systematic temperature dependent Raman studies are extremely important to understand the lattice anharmonicity, phase transition and spin-phonon coupling in bulk and nano materials. The knowledge of temperature coefficients of Raman modes in layered material are useful in estimating the thermal conductivity of these materials \cite{sahoo_temperature-dependent_2013, chen_thermal_2011}. Recently, Kim \emph{et al.} \cite{kim_temperature_2012} have reported the temperature dependence of $A^2_g$ phonon mode in Bi$_2$Se$_3$ single crystal. However, to the best of our knowledge, the temperature dependence of other phonon modes in Bi$_2$Se$_3$ have not been investigated so far. Here, we report a comprehensive temperature dependent Raman study of phonon of different symmetries in Bi$_2$Se$_3$ single crystals grown by modified Bridgman technique. A detailed analysis of the variation of the peak positions and full width at half maximum with respect to the temperature has also been discussed.

\section{Experimental Procedure}
Single crystals of Bi$_2$Se$_3$ were grown using modified Bridgman technique; a high purity ingot(i.e. $5$ N purity) of bismuth and selenium were taken in a stoichiometric ratio in a tapered ampoule and was sealed the tube at a pressure of about $10^{-6}$ Torr. The ampoules were then placed vertically inside a programmable furnace and heated to $850^ o$C for $36$ hours. The temperature of the furnace was then gradually decreased over a period of $5$ to $6$ days to $650^ o$C. Finally, the tubes were quenched in liquid nitrogen. Powder x-ray diffraction (XRD) was used to determine the phase(s) present and the crystal orientation was determined using back scattered Laue diffraction. The samples were also examined using high resolution transmission electron microscopy (HRTEM from JEOL JEM $2200$FS) to ascertain the detailed structure of the crystals. The exfoliated nanoplatelets obtained from Bi$_2$Se$_3$ single crystals were transferred to conventional copper TEM grids for these measurements. Both the room temperature and temperature dependent Raman measurements were carried out using a Horiba-Yobin T$64000$ micro-Raman system and a $532$ \emph{nm} wavelength radiation from a diode laser as an excitation source.

\section{Result and Discussion}
\subsection {X-ray diffraction}
\begin{figure}[!h]
\begin{center}
\includegraphics[width=85mm]{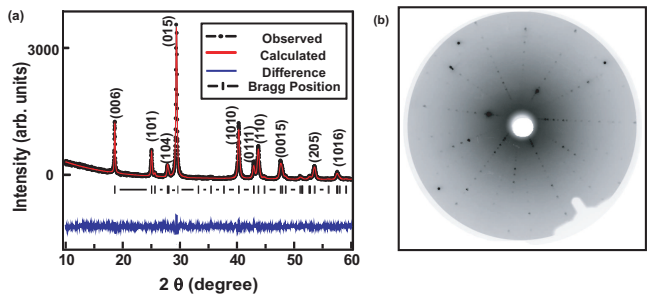}
\caption{(a) Powder XRD along with Rietvald analysis of Bi$_2$Se$_3$ single crystals. (b) Laue diffraction pattern confirming the high quality single crystals.}
\label{fig:figure1}
\end{center}
\end{figure}

Small pieces of single crystals were crushed in to a fine powder for powder x-ray diffraction measurement. A Rietvald analysis were performed and good fit of the experimental data confirms the space group R$\bar{3}$m as shown in Fig. \ref{fig:figure1}(a) and the estimated lattice constants are \emph{a} = $0.42$ \emph{nm} and \emph{c} = $2.87$ \emph{nm}. Back scattered Laue diffraction data show that the crystal are grown along [$001$] as shown in Fig. \ref{fig:figure1}(b).

\subsection {High resolution transmission electron microscopy}
\begin{figure}[!h]
\begin{center}
\includegraphics[width=85mm]{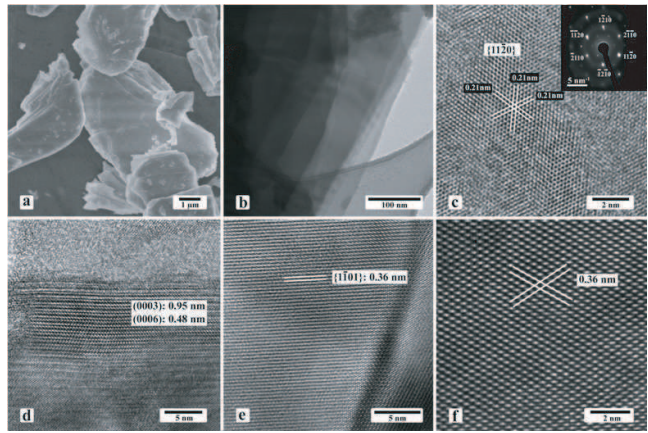}
\caption{(a) SEM image recorded from the synthesized Bi2Se3  material. (b) TEM image showing several individual Bi$_2$Se$_3$ sheets. (c) HRTEM image and SAED pattern (inset). The patterns were indexed to rhombohedral Bi$_2$Se$_3$. (d)  Edge-on view of stack of several a Bi$_2$Se$_3$. (e and f) HRTEM images showing the {1$\bar{1}$01} lattice planes.}
\label{fig:figure2}
\end{center}
\end{figure}

Typical scanning electron microscopy (SEM) and transmission electron microscopy (TEM) images are shown in Fig. \ref{fig:figure2}(a) and \ref{fig:figure2}(b) respectively. The layered structure of Bi$_2$Se$_3$ can be seen in SEM image. The material was easily exfoliated under mild sonication as illustrated in the TEM image. High resolution TEM images of these Bi$_2$Se$_3$ sheets are displayed in Figs. \ref{fig:figure2}(c), \ref{fig:figure2}(d), \ref{fig:figure2}(e) and \ref{fig:figure2}(f).

The single crystalline nature of each Bi$_2$Se$_3$ sheet is demonstrated and shown in Fig. \ref{fig:figure2}(c), the measured lattice spacing ($0.21$ \emph{nm}) agreed well with the {11$\bar{2}$0} spacing (JCPDS card No. $33-0214$). The inset shows selected area electron diffraction pattern (SAED), indexed to rhombohedral Bi$_2$Se$_3$. Also, edge-on image in Fig. \ref{fig:figure2}(d) is shown, with a stack of several Bi$_2$Se$_3$ sheets.  The {$0003$} ($0.95$ \emph{nm}) and {$0006$} ($0.48$ \emph{nm}) lattice planes can be easily identified. Image shown in Fig. \ref{fig:figure2}(e) and \ref{fig:figure2}(f), where the Bi$_2$Se$_3$ {$1\bar{1}01$} lattice planes ($\sim 0.36$ \emph{nm}) are visible.

\subsection {Raman studies}
The atomic arrangement of Bi$_2$Se$_3$ is different from the other members of the layered material family such as graphene and h-BN. Each repeating unit in the crystal is quintuple layer of Se-Bi-Se-Bi-Se which are held together by weak Van der Waals forces. The rhombohedral crystal structure of Bi$_2$Se$_3$ belongs to the $D^5_{3d}$ (R$\bar{3}$m) space group symmetry and there are five atoms per unit cell. Thus, one can have $15$ zone centre phonon branches in the phonon dispersion relation; three acoustic and $12$ optical phonon. Group theory predicts that out of these $12$ zone centre optical phonon $4$ are Raman active and $4$ are infra red (IR) active. The irreducible representations for the zone-center phonon can be written as \cite{kohler_optically_1974}
\begin{equation}\label{1}
\Gamma= 2E_g+ 2A_{1g}+2E_u+2A_{1u}
\end{equation}

\begin{figure}[!h]
\begin{center}
\includegraphics[width=85mm]{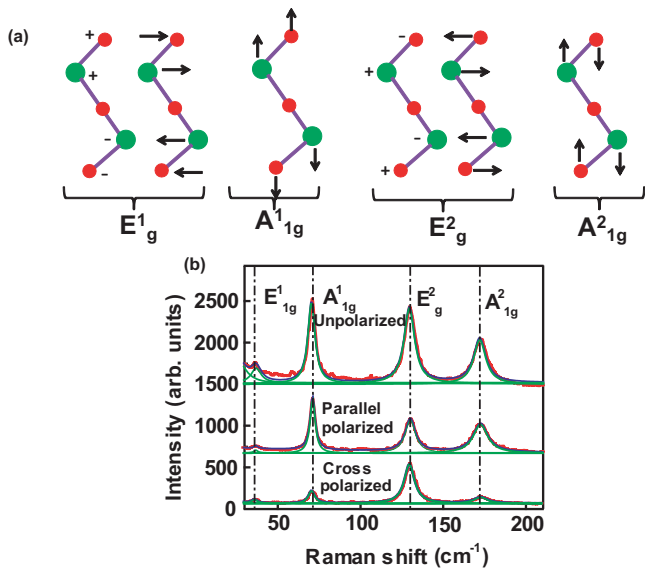}
\caption{(a) Schematic representation of  Raman active modes in rhombohedral compounds of Bi$_2$Se$_3$. (b) Raman spectra with parallel (YY) and perpendicular (YX) polarization configuration of the incident and the scattered light. Red line is the data and blue is the line generated using the fitted peaks in green.}
\label{fig:figure3}
\end{center}
\end{figure}

Note that $E_g$ and $A_{1g}$ modes are Raman active phonon whereas $E_u$ and $A_{1u}$ are IR active modes. There are very few works in literature \cite{zhang_raman_2011} where all four Raman modes have been observed for Bi$_2$Se$_3$. The relative motion of the various atoms for these three observed phonon is schematically presented in Fig. \ref{fig:figure3}(a)

It has been reported that the frequency of vibrations of phonon modes depend on the number of layers and this has been routinely used as a method to identify number of layers in layered materials. It may be noted that the thickness dependent vibrational frequency of $E^1_{g}$ mode is mainly due to the weak Van der Waals inter layer interaction. This mode completely vanishes for single layer graphene \cite{tan_shear_2012} and MoS$_2$ \cite{zeng_low-frequency_2012}.

The non-vanishing polarizability tensor \cite{richter_raman_1977} for these Raman active modes can be given by,

\begin{equation}\label{2}
  A_{g}=\left( \begin{array}{ccc}
    a & 0 & 0 \\
    0 & a & 0 \\
    0 & 0 & a  \end{array}\right),E_{g}=\left( \begin{array}{ccc}
    c & 0 & 0 \\
    0 & -c & d \\
    d & 0 & 0  \end{array}\right),\left( \begin{array}{ccc}
    0 & -c & d \\
    -c & 0 & 0 \\
    d & 0 & 0  \end{array}\right)
\end{equation}
From above polarizability tensors (i.e. Eq. (\ref{2})) it can be easily noticed that for a back scattering geometry with cross polarization configuration (x-y) the $A_{g}$ has vanishing components in the off-diagonal whereas on the other hand $E_g$ mode has non-vanishing off-diagonal elements.

We carried out the polarized Raman scattering on Bi$_2$Se$_3$ single crystals in order to understand the polarization of different phonon. In unpolarized configuration four distinct Raman peaks can be seen at ~ $37$, $72$, $131$ and $175$ cm$^{-1}$ (as shown in Fig. \ref{fig:figure3}(b)) corresponding to $E^1_g$, $A^1_{1g}$, $E^2_g$ and $A^2_{1g}$ respectively, which shows the crystal is in $\alpha$ phase as predicted by group theory \cite{richter_raman_1977, kohler_optically_1974}. Note that the low frequency $E^1_g$ mode has very low intensity and has not been observed in most of the reports available in literature \cite{vilaplana_structural_2011, kim_temperature_2012, gomis_lattice_2011, Zhao_fabrication_2011} possibly due to the high Rayleigh background. If one compares the Raman spectra of parallel and cross polarized states it can be found that the intensities of $A^1_g$ and $A^2_g$ modes have reduced considerably (more than $75$\%) for the cross-polarized state and in contrast, the intensities of E$_g$ modes remain unaltered. These results are consistent with the Raman selection rules discussed in Eq. (\ref{1}). In case of polycrystalline sample it is not possible to observe the polarization effect of $A_{1g}$ mode which is due to the random orientation of sample with respect to the laser polarization. The strong polarization effect of $A^1_g$ and $A^2_g$ modes in our sample indicates good quality single crystals.

\begin{figure}[!h]
\begin{center}
\includegraphics[width=85mm]{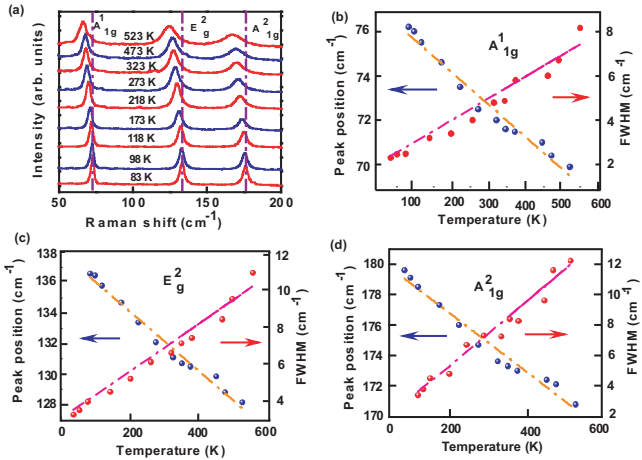}
\caption{(a) Experimental Raman spectra of $\alpha$-Bi$_2$Se$_3$ in the temperature range $83$ K to $523$ K. (b), (c), (d) peak positions and FWHM of different Raman active modes ($A^1_{1g}$, $E^2_g$ and $A^2_{1g}$) recorded with temperature.}
\label{fig:figure4}
\end{center}
\end{figure}

The temperature dependence of the vibrational modes in Bi$_2$Se$_3$ is important for understanding the phonon behavior. Apart from recent work by Kim \emph{et al}. \cite{kim_temperature_2012} there is not much detail available in literature on temperature dependent Raman studies of phonon of different symmetries in Bi$_2$Se$_3$. We carried out a detailed and systematic temperature dependent Raman study on Bi$_2$Se$_3$ in the temperature range $83$ \emph{K} $\le T \le 523$ \emph{K}. Figure \ref{fig:figure4}(a) shows the temperature dependent Raman spectra of Bi$_2$Se$_3$.  As can be seen, the intensities of all Raman active modes (except the $E_{1g}$) are high enough to study the temperature effect on both peak position as well as full width at half maximum (FWHM) over the entire temperature range. With increase in temperature the peak position of $E^2_{g}$, $A^1_{g}$ and $A^2_{g}$ modes were observed to follow systematic red-shift.  Moreover, the FWHM increases with increase in temperature. Temperature dependent broadening of $A^1_g$ mode will be discussed later. The peak positions as well as FWHM of these phonon modes at each temperature were determined by fitting the Raman spectra with a theoretically obtained Raman line shape (damped harmonic oscillator model) which can be written as \cite{sahoo_raman_2012}
\begin{equation}\label{3}
  I(\omega)= \frac{\chi_o\Gamma_o \omega {\omega^2_0}(\bar{n}+1)}{({\omega^2_o}-\omega^2)^2+\omega^2 {\Gamma^2_o}}
\end{equation}

Where
n=exp[$\frac{\hbar\omega}{K_B T}$]-1
is the phonon occupation number, $\omega_o$ and $\Gamma_o$ are the peak position and the line width respectively and $\chi_o$ is related with peak intensity. The change in peak frequency of $A^1_{1g}$, $E^2_g$ and $A^2_{1g}$ modes with increase in the temperature are plotted in Figs. \ref{fig:figure4}(b), \ref{fig:figure4}(c) and \ref{fig:figure4}(d) respectively. One can see from these plots that the peak frequencies vary monotonically with increase in temperature. The following linear equation is used to fit the data \cite{sahoo_temperature-dependent_2013}
\begin{equation}\label{4}
\omega(T)= \omega_o+ \chi T
\end{equation}

where, $\omega_o$ is the frequency of vibration of $E^1_{g}$ or $A_g$ modes at absolute zero temperature, $\chi$ is the first order temperature co-efficient of the $E^1_{2g}$ or $A_g$ modes. The values of temperature coefficient for each Raman mode are listed in Table \ref{table:1}. During these analysis higher order temperature coefficient are not considered as the temperature range used for our experiments is $T \le 523$ K.

\begin{center}
\begin{table}[ht]
\caption{Comparison of Raman peak position, FWHM and temperature co-efficient for different Raman modes in single crystal Bi$_2$Se$_3$. }
\begin{tabular}{c  c  c  c}
  \hline
  Raman mode   &   Peak position  &   FWHM &     $\chi$ \\
   & (cm$^{-1}$) & (cm$^{-1}$) & ($10^{-2} cm^{-1}/K$) \\
   \hline
  $A^1_{1g}$ &  72 & 4.5 & -1.44 \\
  $E^2_{2g}$ & 131 & 6 & -1.95 \\
  $A^2_{1g}$ & 175 & 6.5 & -1.94 \\
  \hline
\end{tabular}
\label{table:1}
\end{table}
\end{center}
The absolute value of the first order temperature coefficient is maximum for $A^2_{g}$ and minimum for $A^1_{g}$. The temperature dependence of vibrational modes is different for different materials, more importantly for a given material it may differ for different phonon modes. Usually, for the first order Raman mode the variation in peak position is mainly due to the volume contribution or thermal expansion that results in anharmonicity. In general phonon frequency $\omega$ is related to the volume and temperature by the following equation \cite{peercy_pressure_1973}
\begin{equation}\label{5}
\begin{split}
\left(\frac{\partial ln\omega}{\partial T}\right)_P & =\left(\frac{\partial lnV}{\partial T}\right)_P\left(\frac{\partial ln\omega}{\partial lnV}\right)_T + \left(\frac{\partial ln\omega}{\partial T}\right)_V \\
& = -\frac{\gamma}{k}\left(\frac{\partial ln\omega}{\partial P}\right)_T + \left(\frac{\partial ln\omega}{\partial T}\right)_V
\end{split}
\end{equation}

where
$\gamma$ $\sim$ $\left(\frac{\partial lnV}{\partial T}\right)_P$ and $k \sim$ -$\left(\frac{\partial ln V}{\partial P}\right)_T$
are respectively the volume thermal coefficient and isothermal volume compressibility . The first term on the right hand side of Eq. (\ref{5}) represents the volume contribution at constant temperature and the second term represents the temperature contribution at constant volume. Hence, anharmonic (pure-temperature) contribution can be determined from the values of $\gamma, k$ and measurable isobaric temperature and isothermal pressure derivative of phonon frequency of the normal modes in Bi$_2$Se$_3$.

A discussion on temperature dependent broadening of A$_{1g}$ mode is in order. Figures \ref{fig:figure4}(b), \ref{fig:figure4}(c) and \ref{fig:figure4}(d) show the plot of full width at half maxima (FWHM) of $A^1_{1g}$, $E^2_g$ and $A^2_{1g}$ modes as a function of temperature. An infinitesimal Raman line width is expected for a perfect harmonic crystal. However, the anharmonicity that exists in the lattice forces allows optical phonons to interchange their vibrational energies to other phonon modes. In that case the optical phonon can decay into either two acoustic phonons of equal and opposite momentum or one optical and one acoustic phonon. The former and latter type of decay are called Klemens and Ridley channel, respectively. It may be pointed out that the phonon lifetime is inversely proportional to the FWHM of the Raman peak. As we have seen from Figs. \ref{fig:figure4}(b), \ref{fig:figure4}(c) and \ref{fig:figure4}(d) that with increase in temperature the FWHM of various phonon modes increase; indicating shorter phonon lifetime at high temperature. In other words the relaxation time of the optical phonon is related to the rate at which it approaches equilibrium. To explain the temperature dependent phonon line width of bismuth selenide the knowledge of phonon dispersion relation and hence the phonon density of state is necessary.  In Bi$_2$Se$_3$ single crystal with several phonon branches, a very complex phonon density of state is expected. In view of this, in the present study it may not be possible to estimate a detail quantitative analysis of the temperature dependent Raman line shape. Thus we will rather focus on a qualitative description of the first order Raman scattering with temperature by generalized decay channel. For convenience we will also consider the only contribution to line width that arises from the decay of zone center optical phonon into one acoustic and one optical phonon. With these considerations, it may be possible to express the temperature dependent phonon line width as \cite{menendez_temperature_1984}

\begin{equation}\label{6}
\Gamma(T)=\Gamma_o + A[1+n(\omega_1, T) + n(\omega_2, T)]
\end{equation}
where, $\Gamma_o$ is the background contribution, A is the anharmonic coefficient, and n($\omega$, T)is the Bose-Einstein distribution function.  The fitted parameters of line width that is used to fit different phonon are listed in Table \ref{table:2}. It may be further noted that anharmonic process that accounts for multi phonon recombination is not the only cause of Raman line shape. Impurities, defects and isotopes can affect the Raman line shape by disturbing the translation symmetry of crystal.

\begin{center}
\begin{table}[h]
\caption{Normal Raman modes of Bi$_2$Se$_3$ with change in background contribution, anharmonic co-efficient, optical and acoustic phonon.}
\begin{tabular}{c   c   c   c   c}
  \hline
  Raman Mode & $\Gamma_o(cm^{-1})$ & $A(cm^{-1})$ & $\omega_1(cm^{-1})$ & $\omega_2(cm^{-1})$ \\
  \hline
  $A^1_{1g}$ & 1.2 & 0.22 & 30 & 24 \\
  $E^2_{2g}$ & 1.6 & 0.60 & 70 & 45 \\
  $A^2_{1g}$ & 1.0 & 0.90 & 80 & 55 \\
  \hline
\end{tabular}
\label{table:2}
\end{table}
\end{center}

\section{Conclusion}
In summary, from the polarized Raman scattering measurement it was found that the $A^1_g$ and $A^2_g$ modes show strong polarization effect, indicating the good quality of single crystals of Bi$_2$Se$_3$ grown using the modified Bridgeman technique. HRTEM and Laue diffraction pattern confirm the high quality single crystals of Bi$_2$Se$_3$. The four Raman modes of the Bi$_2$Se$_3$ in $\alpha$ phase as predicted by group theory were observed in polarized Raman spectra. The temperature dependence of the vibrational modes in Bi$_2$Se$_3$ are found to be almost linear and FWHM of different Raman modes were observed to increases with increase in temperature. The temperature co-efficient for $A^1_{1g}$, $E^2_g$ and $A^2_{1g}$ modes were estimated to be $-1.44\times10^{-2}$, $-1.94\times10^{-2}$ and $-1.95\times10^{-2}cm^{-1}/K$ respectively.

{\bf Acknowledgement}

This work was supported by DOE (grant DE-FG02-ER46526). A P. S. G. thanks NSF fellowship (grant NSF-RII-$1002410$). B.I would like to thank TIFR, Mumbai for providing the facility for crystal growth.



\end{document}